# Defects and hyperfine interactions in binary Fe-Al alloys studied by positron annihilation and Mössbauer spectroscopies[*]

DENG Wen(邓文)[1)], SUN Xiao-Xiang(孙小香), TAN Shao-Xi(谭少希),
LI Yu-Xia(李玉霞), XIONG Ding-Kang(熊定康), HUANG Yu-Yang(黄宇阳)
College of Physical Science and Engineering, Guangxi University, Nanning 530004, China

**Abstract**

The defects, the behavior of 3d electrons and the hyperfine interactions in binary Fe-Al alloys with different Al contents have been studied by the measurements of positron lifetime spectra, coincidence Doppler broadening spectra of positron annihilation radiation and Mössbauer spectra. The results show that on increasing the Al content in Fe-Al alloys, the mean positron lifetime of the alloys increase, while the mean electron density of the alloys decrease. The increase of Al content in binary Fe-Al alloys will decrease the amount of unpaired 3d electrons; as a consequence the probability of positron annihilation with 3d electrons and the hyperfine field decrease rapidly. Mössbauer spectra of binary Fe-Al alloys with Al content less than 25at% show discrete sextets, these alloys give ferromagnetic contribution at room temperature. The Mössbauer spectrum of $Fe_{70}Al_{30}$ shows a broad singlet. As Al content higher than 40 at%, the Mössbauer spectra of these alloys are singlet, that is, the alloys are paramagnetic. The behavior of 3d electron and its effect on the hyperfine field of the binary Fe-Al alloy has been discussed.

## 1  Introduction

Fe-Al alloys are of interest due to their particular mechanical, electrical and magnetic properties [1]. The major obstacle in practical used is in its poor ductility at room temperature. The binary Fe-Al with more than 40 at.% Al content showed mainly intergranular fracture [2]. The room temperature tensile ductility of polycrystalline binary Fe-Al alloy decreases with the increase of Al content [3]. Furthermore, the concentration of constitutional point defect in binary Fe-Al alloy is related to its chemical composition [4].

With the variation of composition and heat treatment, a series of Fe-Al alloys having different soft magnetic and physical properties can be obtained. Small changes in the Al content of the alloy induce large changes in its magnetism [5-7].

The magnetic properties of the alloys are related to their microstructure types. In

Received XX March 2013, Revised XX July 2013

[*] Supported by the National Natural Science Foundation of China (10764001, 51061002) and Program for Science and Technology Innovation Team of Guangxi University

[1)] E-mail: wdeng@gxu.edu.cn

the Fe rich side of the phase diagram the Fe-Al system has a range of disordered A2 bcc structures up to 22 at.% Al at room temperature; on increasing the Al contents, the phase diagram has a variety of intermetallic phases, the first stable compound is $Fe_3Al$ with $DO_3$ crystal structure and it exist over the 18-37 at.% Al range. The other stable compound is FeAl with B2 structure and it exists over the range 37-50 at.% Al [3].

It has been known that binary Fe-Al alloys are of the ferromagnetic disordered type (bcc, A2) for Al contents less than 22 at%, the Fe–25% Al alloy has a large magnetostriction, the Fe–28% Al alloy has a high magnetic permeability [8,9], and the alloy with Al content larger than 35 at% is paramagnetic at room temperature [10]. The magnetic moment per Fe atom in Fe-Al alloy decreases rapidly from 2.0 to 0.7 $\mu_B$ for the Al content ranging from 30 up to 50 at%. At the same time, the average hyperfine field is drastically reduced [11].

The electronic structure of Fe-Al alloy plays an important role in the material's properties, and it has been calculated using a variety of computational methods [12-15]. The relationship between structural and magnetic properties in Fe-Al based magnetic systems has been extensively studied [16-24]. However, experimental studies of the electronic structure and hyperfine interaction in Fe-Al alloy are very limited.

Positron annihilation techniques are well-established to detect open volume and negatively charged centers in solids [25]. Coincidence Doppler broadening spectrum presents information about the one-dimensional momentum distribution of the annihilating positron-electron pair. The 3$d$ electron signal of metals and alloys can be extracted from the coincidence Doppler broadening spectra [26-28]. $^{57}$Fe Mössbauer spectroscopy offers a sensitive microscopic point-probe to identify the nature of the $^{57}$Fe-atom configurations responsible for different hyperfine fields in iron-based alloys [10,11].

In this work, we prepared binary $Fe_{95}Al_5$, $Fe_{90}Al_{10}$, $Fe_{80}Al_{20}$, $Fe_{75}Al_{25}$, $Fe_{70}Al_{30}$, $Fe_{60}Al_{40}$, $Fe_{50}Al_{50}$ and $Fe_{48}Al_{52}$ alloys, α-Fe thin foil, single crystals of Si, Al and Fe. Positron lifetime , coincidence Doppler broadening and Mössbauer spectra of all these samples have been measured. The aim of this work is to analyze the variation of 3$d$ electrons and defects when Fe and Al atoms are combined to form FeAl alloys, to study the behavior of 3d electrons and hyperfine interactions in binary Fe-Al alloys.

## 2  Experimental

The alloy ingots with nominal compositions $Fe_{95}Al_5$, $Fe_{90}Al_{10}$, $Fe_{80}Al_{20}$, $Fe_{75}Al_{25}$, $Fe_{70}Al_{30}$, $Fe_{60}Al_{40}$, $Fe_{50}Al_{50}$ and $Fe_{48}Al_{52}$ alloys were prepared by non-consumable tungsten electrode arc melting in argon atmosphere, using pure Fe and Al. The purity of raw materials used in this study was 99.99 wt.% Fe, 99.98% wt.% Al. All ingots were repeatedly melted more than 3 times by turning over to obtain chemical homogeneity. The ingots were homogenized for 12 h at 1000 °C and then furnace cooled. Alloy specimens, with the thicknesses of 1 mm (for positron annihilation experiments) and 0.2 mm (for Mössbauer experiments), were spark eroded. The surfaces of the specimens were polished. For comparison, α-Fe film, single crystals of Fe, Al and Si were also prepared. After cutting and polishing, the specimens were annealed again at different temperatures (the $Fe_{95}Al_5$, $Fe_{90}Al_{10}$, $Fe_{80}Al_{20}$, $Fe_{75}Al_{25}$, $Fe_{70}Al_{30}$, $Fe_{60}Al_{40}$, $Fe_{50}Al_{50}$ and $Fe_{48}Al_{52}$ alloys, the single crystals of Si and Fe samples at 900 °C, the single crystal of Al sample at 500 °C) for 2 hours in vacuum furnace with a pressure of about $5 \times 10^{-7}$ mbar, and then furnace cooled.

The specimens were subjected to measure the Mössbauer, positron lifetime and coincidence Doppler broadening spectra at room temperature.

Positron lifetime spectra were measured with a fast-fast coincidence ORTEC system. Doppler broadening spectra were measured by using a two-detector coincidence system. $^{22}Na$ source of approximately $3.7 \times 10^5$ Bq was encapsulated in kapton foil, and the source was sandwiched between two identical sample pieces. Mössbauer spectra were measured using a constant-acceleration transducer arranged in a transmission geometry, with a $^{57}Co$ source of approximately $1.85 \times 10^8$ Bq in Rh matrix. The velocity scale for Mössbauer spectra was calibrated related to α-Fe at room temperature.

## 3 Results and Discussion
### 3.1 Defects in binary Fe-Al alloys

The variation of mean positron lifetime ($\tau_m$) with the Al content in binary Fe-Al alloy is shown in Fig. 1. It can be seen that the value of $\tau_m$ increases with Al content in Fe-Al alloys, that is, the average electron density decreases with the Al content in the alloy.

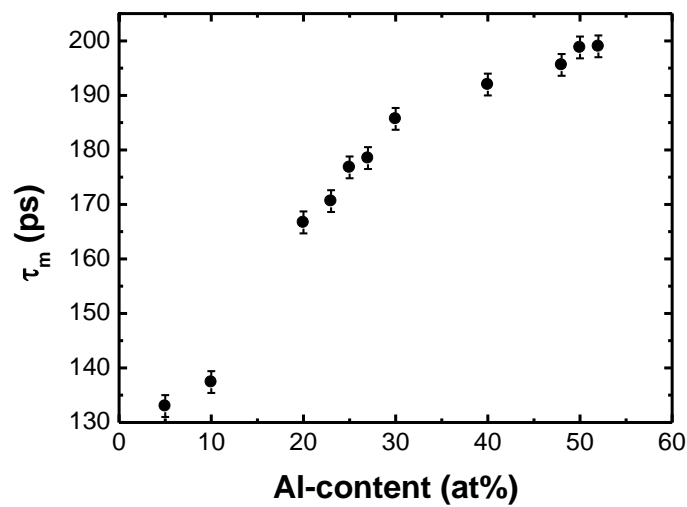

Fig. 1. The variation of mean positron lifetimes ($\tau_m$) with the Al content in binary Fe-Al alloys

As it is known that the positron lifetime in a mono-vacancies of Fe is about 150ps [29,30]. It can be seen in Fig.1, with Al content in Fe-Al alloys increasing from 5 to 10 at.%, $\tau_m$ increases from 133±2 to 137±2 ps, corresponds to less and small open volume defects, i. e. mainly mono-vacancies present in the Fe-Al alloys with disordered A2 phase. With Al content in Fe-Al alloys increasing from 10 to 20 at.%, $\tau_m$ increases from 137±2 to 167±2 ps, it means that there are triple defects with open volume larger than a mono-vacancy occur in the $Fe_{80}Al_{20}$ alloy, these defects may be induced by the $DO_3$ phase present in the alloy (the $DO_3$ phase exist over the 18-33 at.% Al range). The value of $\tau_m$ for $Fe_{60}Al_{40}$ alloy is relative high ($\tau_m$ ($Fe_{60}Al_{40}$) =192±2 ps), it can be due to the defects with large open volume present in the ordered B2 phase. With increasing Al content from 40 at.% to 50 at.%, the alloys contain more and more large open volume defects. It is in agreement with the results reported by Wolff [31] and Gialanella [32].

### 3.2 The 3d electron signals in binary Fe-Al alloys

The use of the coincidence technique reduces the background of the Doppler broadening spectrum of positron annihilation radiation remarkably, and a peak to background ratio of more than $10^5$ can be obtained.

When energetic positron from a $^{22}$Na radioactive source are injected into a solid they slow down to thermal energies, after living in thermal equilibrium, the positron annihilates with an electron from the surrounding medium dominantly into two 511keV gamma quanta. The 511 keV line is Doppler-broadened ($511 \pm \Delta E$) due to the longitudinal momentum $p_L$ component of the annihilating positron-electron pair. $p_L$ is correlated to the Doppler shift $\Delta E$ by $p_L=2\,\Delta E/c$, where $c$ is the light velocity.

The change in shape of the Doppler broadening spectrum due to annihilation with core electrons is very small. To observe the differences among different spectra, we have constructed ratio curves, dividing every spectrum by the spectrum of a reference specimen (the reference sample is a single crystal of Cz-Si in present work) [33]. Before the ratio is taken, all of the spectra have been normalized to a total area of $10^6$ from 511 to 530 keV and a smoothing routine on 9 points was applied. Fig. 2 shows the ratio curves for $Fe_{90}Al_{10}$, $Fe_{75}Al_{25}$, $Fe_{70}Al_{30}$, $Fe_{60}Al_{40}$, $Fe_{50}Al_{50}$ alloys, single crystals of Fe and Al. The abscissa is the longitudinal momentum $p_L$ component of the annihilating positron-electron pair in $10^{-3}\,m_0c$.

In Fig. 2, the ratio curve for the single crystal of Fe shows a very high peak at about $12.5 \times 10^{-3}\,m_0c$, and it is due to positron annihilation with $3d$ electrons of Fe atom [27]. The ratio curve for the single crystal of Al is almost flat due to Al atom without $3d$ electron.

Positrons are very sensitive to vacancies. When a positron is trapped in an open volume defect, the electron density and particularly the core electron density around the defect will be reduced with a consequent decrease in the probability of positron annihilation with core electron [25]. This will decrease the height of the peak of the ratio curve. The ratio curve for $Fe_{50}Al_{50}$ alloy is very low, which is very closed to the ratio curve for Al, as shown in Fig.2 . This can be partly due to the fact that positron trapped in vacancies in $Fe_{50}Al_{50}$ alloy.

It is well known that the electronic configuration of Fe and Al are $1s^22s^22p^63s^23p^63d^64s^2$ and $1s^22s^22p^63s^23p^1$, respectively. There are some 3d electrons in a Fe atom without being coupled. When Fe and Al atoms are combined to form Fe-Al alloys, some of the $3d$ electrons of Fe atoms and $3p$ electrons of Al atoms are localized to form strong covalent bonds [12,13] and thus decreasing the probability of positron annihilation with $3d$ electrons of Fe atoms.

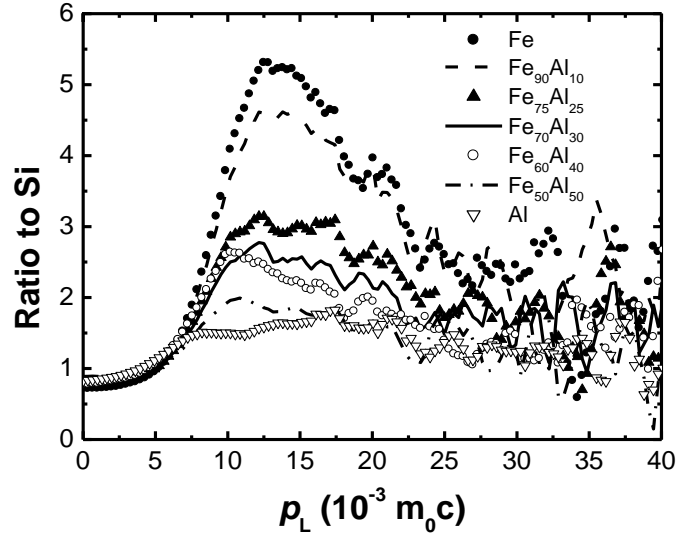

Fig. 2. The ratio curves for $Fe_{90}Al_{10}$, $Fe_{75}Al_{25}$, $Fe_{70}Al_{30}$, $Fe_{60}Al_{40}$, $Fe_{50}Al_{50}$ alloys, single crystals of Fe and Al. The reference sample is a single crystal of Si.

Fig. 2 shows that the $3d$ electron signal in the spectrum of the alloy decreases with the increase of Al content in Fe-Al alloys. It can be seen in Fig. 2, the $3d$ electron signal in the spectrum of $Fe_{90}Al_{10}$ alloy with disordered A2 phase is much higher than that of $Fe_{75}Al_{25}$ alloy with $DO_3$ phase. And the $3d$ electron signal in the spectrum of $Fe_{50}Al_{50}$ alloy with B2 phase is much lower than that of $Fe_{75}Al_{25}$ alloy. This can be due to the following reasons: the concentration of structural vacancies in Fe-Al increases with Al content. On the other hand, the increase of Al content in Fe-Al alloys will enhance the Fe $d$-Al $p$ interactions, and weaken the $d$-$d$ interactions, as a consequence the probability of positron annihilation with $3d$ electrons of Fe atoms decreases rapidly.

The results we obtained from the coincidence Doppler broadening spectrum of positron annihilation radiation is in good agreement with that of the positron lifetime.

### 3.3 Mössbauer spectra of binary Fe-Al alloys

The Mössbauer spectra for α-Fe, $Fe_{90}Al_{10}$, $Fe_{80}Al_{20}$, $Fe_{75}Al_{25}$ $Fe_{73}Al_{27}$, $Fe_{70}Al_{30}$, $Fe_{60}Al_{40}$, and $Fe_{50}Al_{50}$ alloys are shown in Fig 3 (a) and (b). For comparison, we do not make any fit on the data.

The comparison of the Mössbauer spectra of the studied samples indicates that in Fe–Al binary alloys Mössbauer spectra features are very dependent on Al content. As Al content lower than 25 at%, the Mössbauer spectra of these alloys show discrete sextets (see Fig. 3(a) ), that is, these alloys give ferromagnetic contribution at room temperature. As Al content higher than 40 at%, the Mössbauer spectra of these alloys show no sextets at all, they are singlet ( see Fig.3(b) ). The Mössbauer spectrum of $Fe_{70}Al_{30}$ shows blurred ones, there are Fe atoms with magnetic and non-magnetic behaviors, but the contributions observed in the spectrum is not discrete.

The magnetic contributions of Fe atoms are related directly to the presence of the unpaired 3d electrons of the atoms. The increase of Al content in binary Fe-Al alloys will increase the Fe *d*-Al *p* interactions, while decrease the amount of unpaired 3d electrons and the hyperfine field in the alloys. The hyperfine field ($H_{hf}$) is directly proportional to the distant between the first peak position and the sixth peak position of a Mössbauer spectrum.

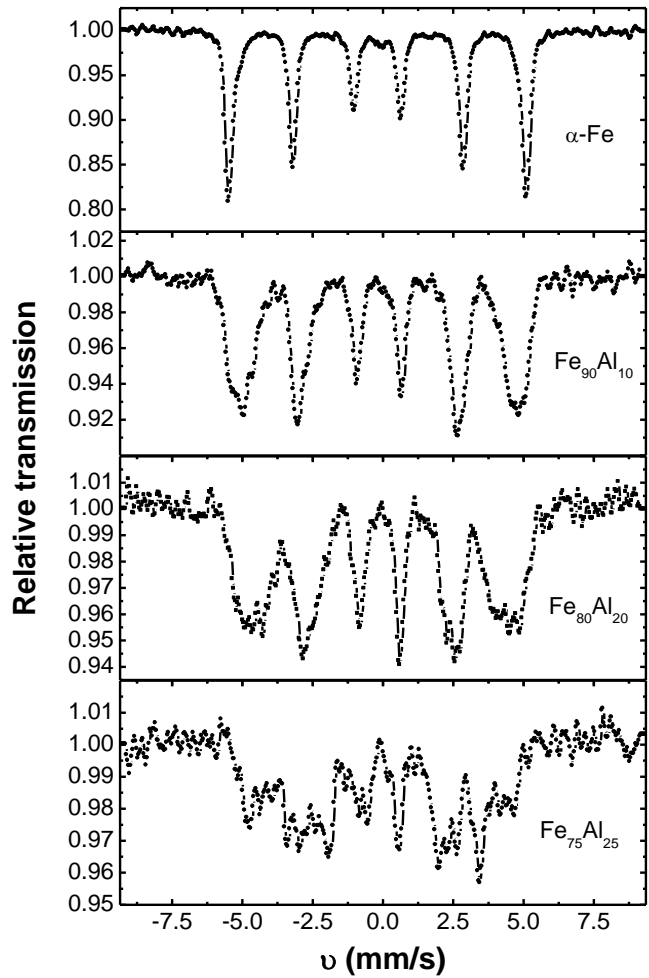

Fig. 3 (a) The Mössbauer spectra for α-Fe, $Fe_{90}Al_{10}$, $Fe_{80}Al_{20}$ and $Fe_{75}Al_{25}$ alloys.

In a pure α-Fe sample, most of the 3d electrons of Fe atoms are unpaired, all Fe atoms are magnetic, corresponds to a higher hyperfine field in the sample (it is known as 330.6 kOe), and its Mössbauer spectrum shows discrete sextet, the relative intensities of the six peaks are in the ratio 3 : 2 : 1 : 1 : 2 : 3 ( see Fig. 3 (a)).

For the binary Fe-Al alloys with Al content lower than 25 at%, some of the 3*d* electrons of Fe atoms and 3*p* electrons of Al atoms are localized to form strong covalent bonds, while the other 3d electrons remain unpaired. The Fe atoms with unpaired 3d electrons are magnetic. The Mössbauer spectrum of these alloys shows discrete sextets, however, the relative intensities of the six peaks and the hyperfine field ($H_{hf}$) decrease with the increase of Al content (see Fig. 3(a)).

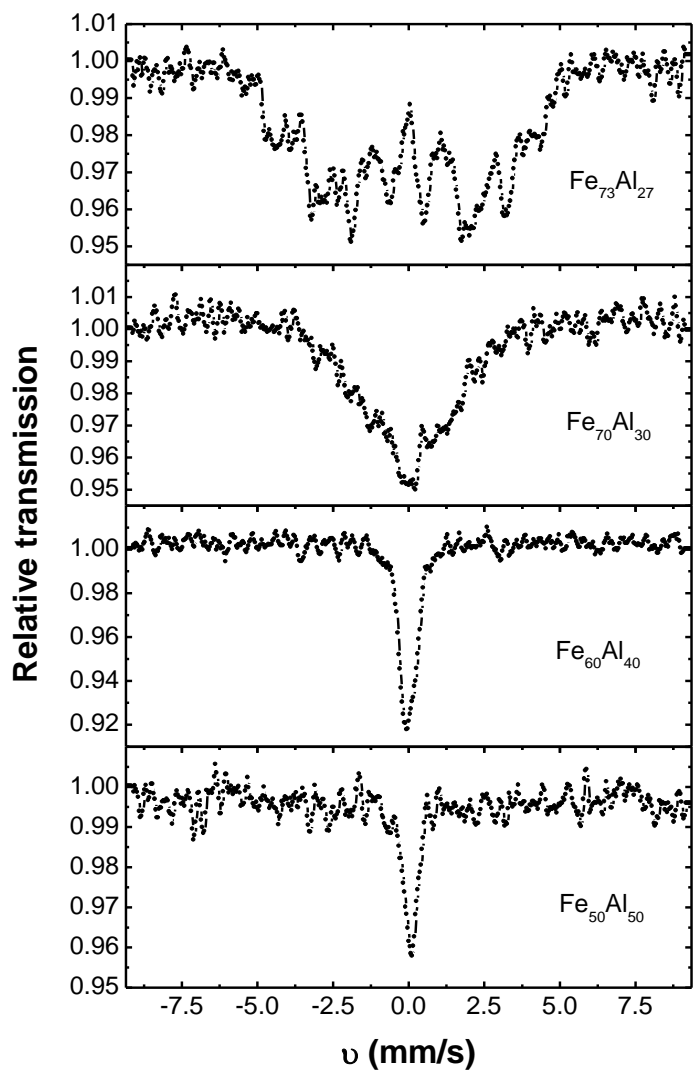

Fig. 3 (b) The Mössbauer spectra for $Fe_{73}Al_{27}$, $Fe_{70}Al_{30}$, $Fe_{60}Al_{40}$, and $Fe_{50}Al_{50}$ alloys.

For the binary Fe-Al alloys with Al content higher than 40 at%, most of the $3d$ electrons of Fe atoms are bonded with $3p$ electrons of Al atoms. There are very few unpaired 3d electron in the alloys, the alloys are paramagnetic, and the Mössbauer spectra of these alloys are singlet ( see Fig.3(b) ).

$Fe_{70}Al_{30}$ alloy shows a very different Mössbauer spectrum, where the discrete sextets disappear and a paramagnetic singlet appears, superposed to a blurred magnetic distribution. The Mössbauer spectrum of $Fe_{70}Al_{30}$ shows a broad singlet ( see Fig.3(b) ).

The results we obtained from the Mössbauer experiment is in good agreement with that of coincidence Doppler broadening spectra.

## 4 Conclusions

(1) The increase of Al content in binary Fe-Al alloys will enhance the 3d-3p interactions and decrease the amount of unpaired 3d electrons, as a consequence the probability of positron annihilation with the 3d electrons of Fe atoms and the hyperfine field decrease rapidly.

(2) The concentration of structure vacancies in binary Fe-Al alloy increases with Al content.

(3) Mössbauer spectra of binary Fe-Al alloys with Al content less than 25at% show discrete sextets, these alloys give ferromagnetic contribution at room temperature. The Mössbauer spectrum of $Fe_{70}Al_{30}$ shows a broad singlet. As Al content higher than 40 at%, the Mössbauer spectra of these alloys show singlet, and the alloys are paramagnetic.